\documentstyle [12pt,a4]{article}

\newcommand{\be}{\begin{equation}}
\newcommand{\ee}{\end{equation}}
\newcommand{\bea}{\begin{eqnarray}}
\newcommand{\ena}{\end{eqnarray}}
\newcommand{\sect}[1]{\setcounter{equation}{0}\section{#1}}
\newcommand{\vs}[1]{\rule[- #1 mm]{0mm}{#1 mm}}

\newcommand{\sm}[2]{\frac{\mbox{\footnotesize #1}\vs{-2}}
                   {\vs{-2}\mbox{\footnotesize #2}}}

\newcommand{\shalf}{\sm{1}{2}}

\newcommand{\wl}{\widetilde{L}}

\newcommand{\NP}[1]{Nucl.\ Phys.\ {\bf #1}}
\newcommand{\PL}[1]{Phys.\ Lett.\ {\bf #1}}

\newcommand{\CMP}[1]{Comm.\ Math.\ Phys.\ {\bf #1}}
\newcommand{\PR}[1]{Phys.\ Rev.\ {\bf #1}}
\newcommand{\PRL}[1]{Phys.\ Rev.\ Lett.\ {\bf #1}}
\newcommand{\MPL}[1]{Mod.\ Phys.\ Lett.\ {\bf #1}}

\begin{document}
\renewcommand{\thefootnote}{\fnsymbol{footnote}}
\newpage
\setcounter{page}{0}

\vs{20}

\begin{center}
{\LARGE {\bf Generalised $q$-Deformed Oscillators\\[0.5cm]
 and their Statistics}}\\[2cm]

{\large Dao Vong Duc \footnote{On leave of absence from Institute of Physics,
P.O. Box 429 Bo Ho, Hanoi 10000, Vietnam}}\\[.5cm]

{\em {Laboratoire de Physique Th\'eorique}}
{\small E}N{\large S}{\Large L}{\large A}P{\small P}
\footnote{URA 14-36 du CNRS, associ\'ee \`a l'E.N.S. de Lyon.}\\
{\em Chemin de Bellevue, BP 110, F - 74941 Annecy-le-Vieux Cedex,
France}
\end{center}

\vs{10}

\centerline{\bf Abstract}

\vs{2}

{\indent We consider a version of generalised $q$-oscillators and some of
their applications. The generalisation includes also "quons" of infinite
statistics and deformed oscillators of parastatistics. The statistical
distributions for different $q$-oscillators are derived for their
corresponding Fock
space representations. The deformed Virasoro algebra and SU(2) algebra are also
treated.}

\vs{8}

\rightline{{\small E}N{\large S}{\Large L}{\large A}P{\small P}-A-494/94}
\rightline{October 1994}

\newpage

\sect{Introduction}

Over the last few years there has been a growing activity in the study of
Quantum Groups and Algebras \cite{1}-\cite{3}. This is connected with the fact
that
these new mathematical structures are relevant for a variety of diverse
problems in Theoretical Physics, such as quantum inverse scattering theory,
exactly solvable model in statistical mechanics, rational conformal field
theory, two-dimensional field theory with fractional statistics, etc...

The algebraic structure of quantum group can be formally described as a
$q$-deformation of the enveloping algebra $U(G)$ of a Lie algebra $G$ in such a
way that $G$ is recovered in the limit of the deformation parameter $q
\rightarrow 1$. In this connection the concept of $q$-deformed quantum harmonic
oscillators has also been introduced \cite{4,5}, which prove to be powerful for
studying the representation of  $q$-deformed enveloping algebra. In particular
the $q$-formalism has been developed for $SU(2)$ group and some others.

The study of $q$-oscillators has been, on the other hand, stimulated during the
recent years by the increasing interests in particles obeying statistics
different from Bose and Fermi.

In this paper we would like to consider a version of generalised
$q$-oscillators and some of their applications. This generalisation includes on
an equal footing the usual $q$-deformed oscillators \cite{4,5} and the "quons"
of infinite statistics \cite{6}-\cite{9}. We also extend the formalism for the
deformation of parastatistics \cite{10,11}. The contents of the paper are
arranged as follows. In section 2 we consider the $q$-deformed single mode
oscillators, their satatistics, and associated $q$-deformed Virasoro algebra.
In section 3 we are dealing with multimode oscillators and the representations
of ($q$-deformed) $SU(2)$ algebra. Section 4 is devoted to $q$-deformed
paraoscillators.

\sect{Single Mode Oscillator.}

\indent

{\bf 1.} We consider the bosonic oscillator with the deformed commutation rule
of the form:
\be
aa^+ - qa^+a = q^{cN}
\label{eq:2.1}
\ee
where $N$ is oscillator number operator,
\be
[N,a] =-a
\label{eq:2.2}
\ee
$q$ and $c$ are some parameters.

The usual $q$-deformation
\[
aa^+ - qa^+a = q^{-N}
\]
corresponds to the value $c=-1$, and the "infinite statistics"
\be
aa^+ = 1
\label{eq:2.3}
\ee
corresponds to $c=0, q=0$.

Equation (\ref{eq:2.1}) gives:
\be
aa^{+n} = q^n a^{+n} a + [n]^{(c)}_q (a^+)^{n-1} q^{cN}
\label{eq:2.4}
\ee
where the general notation
\be
[x]_q^{(c)} \equiv \frac{q^x-q^{cx}}{q-q^c}
\label{eq:2.5}
\ee
is used.

It is seen from the equation (\ref{eq:2.4}) that the algebra (\ref{eq:2.1}) can
be
realised in the Fock space spanned by the orthonormalised eigenstates of the
operator $N$,
\be
|n> \equiv \frac{1}{\sqrt{[n]^{(c)}_q !}} (a^+)^n|0>
\label{eq:2.6}
\ee
and in this space the following relations hold:
\be
a^+a=[N]^{(c)}_q, \ \ aa^+ = [N+1]^{(c)}_q
\label{eq:2.7}
\ee

For the calculations it is helpful to use the identities:
\bea
{[}x{]}^{(c)}_q &=& [x]^{(c^{-1})}_{qc} \nonumber \\
{[}-x{]}^{(c)}_q &=& -q^{-(c+1)} [x]^{(c)}_{q^{-1}} \nonumber \\
{[}x+y{]}^{(c)}_q &=& q^y[x]^{(c)}_q + q^{cx} [y]^{(c)}_q = q^{cy} [x]^{(c)}_q
+
q^x [y]^{(c)}_q
\label{eq:2.8}
\ena

\vs{3}

{\bf 2.} For deformed fermionic oscillator we put:
\be
bb^+ + qb^+b = q^{cN}, \ \ [N,b] =-b
\label{eq:2.9}
\ee

This algebra can be realised in the Fock space spanned by the orthonormalised
eigenstates of $N$,
\be
|n>= \frac{1}{\sqrt{\{n\}^{(c)}_q !}} \ b^{+n} |0>
\label{eq:2.10}
\ee

Here the notation
\be
\{n\}^{(c)}_q = \frac{q^{cn} + (-1)^{n+1} q^n}{q^c+q}
\label{eq:2.11}
\ee
is used.

Using the equation
\be
bb^{+n} = (-1)^n q^n b^{+n} b + \{ n \}^{(c)}_q (b^+)^{n-1} q^{cN}
\label{eq:2.12}
\ee
it can be shown that in this Fock space the following relations hold:
\be
b^+b = \{ N \}^{(c)}_q, \ bb^+ = \{ N+1 \}^{(c)}_q
\label{eq:2.13}
\ee

{\bf 3.} Consider now the deformed Green function defined as the statistical
distribution of $a^+a$ and $b^+b$. The statistical distribution of the operator
$F$ is defined through the formula:
\be
<F> = \frac{1}{Z} \ Tr \left( e^{-\beta H} F \right)
\label{eq:2.14}
\ee
where $Z$ is the partition function,
\[
Z \equiv Tr (e^{-\beta H})
\]
which determines the thermodynamic properties of the system, $\beta =
\frac{1}{KT}, H$ is Hamiltonian, which is usually taken of the form $H=\omega
N, \ \omega$ being one particle-oscillator energy. The trace must be taken
over a complete set of states.

The calculations give the following results:
\bea
<a^+a> &=& \frac{e^{\beta \omega} -1}{e^{2 \beta \omega} - (q+q^c) e^{\beta
\omega} + q^{1+c}} \label{eq:2.15a} \\
<b^+b> &=& \frac{e^{\beta \omega} -1}{e^{2 \beta \omega} + (q-q^c) e^{\beta
\omega} - q^{1+c}}
\label{eq:2.15b}
\ena

 From these we recover the familiar formulae
\bea
<a^+a> = \frac{1}{e^{\beta \omega}-1} \nonumber \\
<b^+b> = \frac{1}{e^{\beta \omega}+1} \nonumber
\ena
for Bose and Fermi statistics when $q=1$, and the result \cite{12}
\bea
<a^+a> &=& \frac{e^{\beta \omega} -1}{e^{2 \beta \omega} - (q+q^{-1}) e^{\beta
\omega} + 1} \nonumber \\
<b^+b> &=& \frac{e^{\beta \omega} -1}{e^{2 \beta \omega} + (q-q^{-1}) e^{\beta
\omega} +1} \nonumber
\ena
for usual $q$-deformed statistics when $c=-1$.

In the limit of "infinite statistics", $c=0, q=0$, we have:
\be
<a^+a> = <b^+b> = e^{-\beta \omega}
\label{eq:2.16}
\ee

{\bf 4}. The Virasoro algebra plays a crucial role in string theory. The
centreless Virasoro algebra consists of the generators $L_n, n \in Z$,
satisfying the commutation relation:
\be
[L_n, L_m] = (n-m) L_{n+m}
\label{eq:2.17}
\ee

In the (undeformed) oscillator formalism this algebra can be realised by
putting
\be
L_n = (a^+)^{-n+1} a
\label{eq:2.18}
\ee

Consider now the $q$-deformation of Virasoro algebra based on the
$q$-oscillator algebra (\ref{eq:2.1}). A version of such deformation can be
obtained if instead of (\ref{eq:2.18}) we take the generators of the form:
\be
\wl_n = q^{cN} (a^+)^{-n+1} a
\label{eq:2.19}
\ee

We then have:
\be
\wl_n \wl_m - q^{(c+1)(n-m)} \ \ \wl_m \wl_n = q^{c(2N+n+m)} [n-m]^{(c)}_q
\ \ \wl_{n+m}
\label{eq:2.20}
\ee

The usual $q$-deformed Virasoro algebra
\[
[\wl_n, \wl_m]=q^{-(2N+n+m)}[n-m]_q \ \ \wl_{n+m}
\]
is recovered by putting $c=-1$.

For the case of infinite statistics, $c=0, q=0$, equation (\ref{eq:2.20})
gives:
\be
\wl_n \wl_m = \wl_{n+m}
\label{eq:2.21}
\ee

This result can also be checked directly from (\ref{eq:2.3}) and the expression
of $\wl_n$,
\[
\wl_n = (a^+)^{-n+1} a
\]
followed from (\ref{eq:2.19}) with $c=0, q=0$.

\sect{Multimode Oscillator.}

\indent

The formulae (\ref{eq:2.1}) and (\ref{eq:2.9}) for single mode oscillators can
be generalised for multimode oscillators as follows:
\be
A_i A_j^+ \mp \left\{ (q-q^{\delta(c)}) \delta_{ij} + q^{\delta(c)} \right\}
A^+_j A_i = \delta_{ij} \ q^{cN_i}
\label{eq:3.1}
\ee
\be
e^{cA_i} A_j \ e^{-cA_i} = A_j
\label{eq:3.2}
\ee
\be
[N_i, A_j] = - \delta_{ij} \ A_i
\label{eq:3.3}
\ee
where $A$ stands for $a$ or $b$,
\[
\delta(c) =
\left\{
\begin{array}{ccc}
1 & , & c=0 \\
0 & , & c \neq 0
\end{array}
\right.
\]
and the upper (lower) sign in (\ref{eq:3.1}) refers to the bosonic (fermionic)
case.

Equation (\ref{eq:3.2}) means that $[A_i, A_j]=0$ when $c \neq 0$, but no
commutation rule is imposed on $A_iA_j$ when $c = 0$. In the latter case
(\ref{eq:3.1}) reads \cite{12}:
\be
A_i A_j^+ \mp q \ A^+_j A_i = \delta_{ij}
\label{eq:3.4}
\ee

Let us consider now the realisation of (deformed) $SU(2)$ algebra based on the
$q$-oscillator algebra (\ref{eq:3.1})-(\ref{eq:3.3}). In the case $c \neq 0$
this can be performed in the Fock space spanned by the orthonormalised
eigenstates of $N_1$ and $N_2$ defined as
\be
|jm> = \frac{1}{\sqrt{[j-m]^{(c)}_q ! [j+m]^{(c)}_q !}} \ (a_1^+)^{j+m}
(a^+_2)^{j-m} |0>
\label{eq:3.5}
\ee
with the identification:
\bea
E &=& q^{- \shalf (1+c)N_2} {a_1}^+ a_2  \nonumber \\
F &=& q^{- \shalf (1+c)(N_2-1)} {a_2}^+ a_1  \nonumber \\
H &=& N_1 - N_2 \label{eq:3.6}
\ena

In fact, using the identities (\ref{eq:2.8}) it can be shown that
\[
[H,E]=2E, \ \ [H,F] = -2F
\]
\be
EF-q^{-(1+c)} \ \ FE = [H]_q^{(c)}
\label{eq:3.7}
\ee

 From here, in particular, we recover the usual one-parameter deformed
$SU(2)_q$
algebra \cite{4,5} when $c=-1$:
\[
[E,F]=[H]_q \equiv \frac{q^H-q^{-H}}{q-q^{-1}}
\]

In the case of infinite statistics, $c=0, q=0$, the representation of $SU(2)$
can be constructed in the Fock space spanned by the orthonormalised eigenstates
\be
|jm;(r),(s)> = \prod^r_{K=1} (a_1^+)^{r_K} (a^+_2)^{S_K} |0>
\label{eq:3.8}
\ee
\[
\sum^p_{K=1} r_K = j+m, \ \ \ \ \sum^p_{K=1} S_K = j-m
\]
by putting
\bea
E &=& \sum_{i=1,2} \ \ \sum^\infty_{K=0} a^+_{i_K}...a^+_{i_2} a^+_{i_1} .
a_1^+
a_2 . a_{i_1}. a_{i_2} ... a_{i_K} \nonumber \\
F &=& \sum_{i=1,2} \ \ \sum^\infty_{K=0} a^+_{i_K}...a^+_{i_2} a^+_{i_1} .
a_2^+
a_1 . a_{i_1}. a_{i_2} ... a_{i_K} \nonumber \\
H &=& N_1 - N_2 \nonumber \\
N_j &=& \sum_{i=1,2} \ \ \sum^\infty_{K=0} a^+_{i_K}...a^+_{i_2} a^+_{i_1} .
a_j^+
a_j . a_{i_1} . a_{i_2} ... a_{i_K} \label{eq:3.9}
\ena

Note that the dimension of representation with spin $j$ is $2j+1$ for $j=0,
\shalf$, and more than $2j+1$ for $j \geq 1$. Thus, we have two-dimensional
representation for $j=\shalf$ with state vectors:
\[
| \shalf, \shalf>\ =a_1^+|0>, \ \ \ \ \ \ \ |\shalf, -\shalf> = a^+_2|0>,
\]
four-dimensional representation for $j=1$ with state vectors:
\[
|1,1>=(a_1^+)^2 | 0>
\]
\[
|1,0;(1)> = a_1^+ a_2^+|0> , \ \ \ \ |1,0;(2)> = a^+_2 a_1^+ |0>
\]
\[
|1,-1>=(a_2^+)^2|0>
\]
etc.

\sect{Deformed para-Oscillator.}

\indent

Here we restrict the consideration to the case of single mode parabose
oscillator only. As has been shown in \cite{13,14} the (undeformed) parabose
oscillator of order $p$ obeys the commutation relations:
\bea
{[}a,a^+{]} &=& 1 + (-1)^N (p-1) \nonumber \\
{[}N,a{]} &=& -a \nonumber \\
N &=& \shalf (a^+ a + aa^+ -p)
\label{eq:4.1}
\ena

In the Fock space spanned by the state vectors $|n> \sim (a^+)^n |0>$,
(\ref{eq:4.1}) gives:
\bea
aa^+ |n  > &=& |n>.
\left\{
\begin{array}{ll}
(n+p), & n \ \ \mbox{even} \\
(n+1), & n \ \ \mbox{odd}
\end{array}
\right. \nonumber \\
a^+a |n  > &=& |n>.
\left\{
\begin{array}{ll}
n, & n \ \ \mbox{even} \\
(n+p-1), & n \ \ \mbox{odd}
\end{array}
\right. \label{eq:4.2}
\ena
For the $q$-deformation we propose:
\bea
aa^+ |n  > &=& |n>.
\left\{
\begin{array}{ll}
{[}n+p{]}^{(c)}_q, & n \ \ \mbox{even} \\
{[}n+1{]}^{(c)}_q, & n \ \ \mbox{odd}
\end{array}
\right. \nonumber \\
a^+a |n  > &=& |n>.
\left\{
\begin{array}{ll}
{[}n{]}^{(c)}_q, & n \ \ \mbox{even} \\
{[}n+p-1{]}^{(c)}_q, & n \ \ \mbox{odd}
\end{array}
\right. \label{eq:4.3}
\ena
as the generalisation of (\ref{eq:4.2}).

This means that in the Fock space of states $|n>$ the following equations
hold:
\bea
aa^+ &=& \shalf (1+(-1)^N) [N+p]^{(c)}_q + \shalf (1-(-1)^N) [N+1]^{(c)}_q
\nonumber \\
a^+a &=& \shalf (1+(-1)^N) [N]^{(c)}_q + \shalf (1-(-1)^N) [N+p-1]^{(c)}_q
\label{eq:4.4}
\ena

 From these, by using the identities (\ref{eq:2.8}), we can derive the
following
commutation relation:
\[aa^+ -q^{1+(p-1)(-1)^N} a^+a
\]
\be
= \left\{ \shalf (1+(-1)^N)[p]^{(c)}_q - \shalf (1-(-1)^N) [p-2]^{(c)}_q
.q^{c+2-p} \right\} q^{cN}
\label{eq:4.5}
\ee

 From (\ref{eq:4.5}) we can obtain the recurrent formula:
\bea
aa^{+n} &=& q^{n+\shalf(1-(-1)^n).(p-1)(-1)^N} a^{+^n}a \nonumber \\
&& + \shalf \sum^{[\frac{n-1}{2}]}_{K=0} q^{2(1-c)K} f(N) \ q^{cN} (a^+)^{n-1}
\label{eq:4.6} \\
&& + \shalf \left( \sum^{[\frac{n}{2}]}_{K=0} q^{2(1-c)K} -1 \right)
q^{c-1+(p-1).(-1)^N} f(N-1)q^{cN} (a^+)^{n-1}
\ena
where
\[
f(N) \equiv (1+(-1)^N) [p]^{(c)}_q -(1-(-1)^N)[p-2]^{(c)}_q. q^{c+2-p}
\]

Equation (\ref{eq:4.6}) allows us to have the following expression for the norm
of the state $(a^+)^n|0>$:
\[
<0|a^na^{+n} |0>
\]
\be
= \shalf \ q^{c(n-1)} \left\{ \sum^{[\frac{n-1}{2}]}_{K=0} q^{2(1-c)K} f(n-1) +
\left( \sum^{[\frac{n}{2}]}_{K=0} q^{2(1-c)K}-1 \right) . \ q^{c-1-(p-1)(-1)^n}
f(n) \right\}
\ee

Finally, let us be interested in the statistical distribution of $a^+a$. The
calculations give the following result:
\be
<a^+a> = \frac{1-e^{-\beta \omega}}{q-q^c} \left\{ \frac{1+q^p e^{-\beta
\omega}}{1-q^2 e^{-2 \beta \omega}} - \frac{1+q^{cp} e^{-\beta
\omega}}{1-q^{2c} e^{-2\beta \omega}} \right\}
\label{eq:4.8}
\ee

It is straighforward to check that when $p=1$ this recovers the formula
(\ref{eq:2.15a}).

\vs{3}

{\large {\bf Acknowledgments}}

\indent

I would like to express my sincere thanks to the members of Laboratoire de
Physique Th\'eorique ENSLAPP for the kind hopitality during my stay at LAPP. I
am particularly grateful to P. Aurenche and P. Sorba for the kind support and
permanent attention. My thanks are also due to D. Poencier for taking care in
typing the manuscript.

\end{document}